\documentclass{IEEEtran}
\usepackage{cite}
\usepackage{amsmath,amssymb,amsfonts}
\usepackage{algorithmic}
\usepackage{graphicx}
\usepackage{textcomp}
\usepackage{capt-of}

\AtBeginDocument{%
  \providecommand\BibTeX{{%
    Bib\TeX}}}

\usepackage[dvipsnames]{xcolor}
\usepackage{url}
\usepackage[hidelinks]{hyperref}
\usepackage{booktabs}
\usepackage{listings}
\usepackage{balance}

\definecolor{darkpastelgreen}{rgb}{0.01, 0.75, 0.24}

\newcommand{\sectionname}{Section}

\usepackage{listings} 
\usepackage{caption} 
\newcommand\codestyle{\lstset{
language=C++,
basicstyle=\ttfamily,
morekeywords={uint32_t,constexpr}, 
keywordstyle=\ttfamily\color{BlueViolet},
emph={s_float,s_float_simd}, 
emphstyle=\ttfamily\color{BrickRed}, 
stringstyle=\color{ForestGreen},
frame=single, 
showstringspaces=false,
commentstyle=\ttfamily,
numbers=left,
stepnumber=1,
literate={-}{-}1,
columns=fullflexible,
numberstyle=\tiny,
}}

\lstnewenvironment{code}[1][]{%
    \codestyle\footnotesize\lstset{#1}}{}
\let\origthelstnumber\thelstnumber
\makeatletter
\newcommand*\Suppressnumber{%
  \lst@AddToHook{OnNewLine}{%
    \let\thelstnumber\relax%
     \advance\c@lstnumber-\@ne\relax%
    }%
}

\newcommand*\Reactivatenumber[1]{%
  \lst@AddToHook{OnNewLine}{%
   \let\thelstnumber\origthelstnumber%
   \setcounter{lstnumber}{\numexpr#1-1\relax}%
  }%
}

\makeatother

\def\BibTeX{{\rm B\kern-.05em{\sc i\kern-.025em b}\kern-.08em
    T\kern-.1667em\lower.7ex\hbox{E}\kern-.125emX}}
\begin{document}

\title{QOPS: A Compiler Framework for Quantum Circuit Simulation Acceleration with Profile Guided Optimizations}

\DeclareRobustCommand{\IEEEauthorrefmark}[1]{\smash{\textsuperscript{\footnotesize #1}}}
\author{
	\IEEEauthorblockN{
    Yu-Tsung Wu\IEEEauthorrefmark{1}\textsuperscript{\textsection},
    Po-Hsuan Huang\IEEEauthorrefmark{1,2}\textsuperscript{\textsection},
    Kai-Chieh Chang\IEEEauthorrefmark{1},
		Chia-Heng Tu\IEEEauthorrefmark{1}, and 
    Shih-Hao Hung\IEEEauthorrefmark{2,3}}\\
	\IEEEauthorblockA{
    \IEEEauthorrefmark{1}National Cheng Kung University, Tainan, Taiwan\\
		\IEEEauthorrefmark{2}National Taiwan University, Taipei, Taiwan\\
    \IEEEauthorrefmark{3}Mohamed bin Zayed University of Artificial Intelligence, Masdar City, Abu Dhabi, United Arab Emirates\\
  }
}

\maketitle
\begingroup\renewcommand\thefootnote{\textsection}
\footnotetext{These authors contributed equally to this work.}
\endgroup

\begin{abstract}
Quantum circuit simulation is important in the evolution of quantum software and hardware. Novel algorithms can be developed and evaluated by performing quantum circuit simulations on classical computers before physical quantum computers are available. Unfortunately, compared with a physical quantum computer, a prolonged simulation time hampers the rapid development of quantum algorithms. Inspired by the feedback-directed optimization scheme used by classical compilers to improve the generated code, this work proposes a quantum compiler framework QOPS to enable profile-guided optimization (PGO) for quantum circuit simulation acceleration. The QOPS compiler instruments a quantum simulator to collect performance data during the circuit simulation and it then generates the optimized version of the quantum circuit based on the collected data. Experimental results show the PGO can effectively shorten the simulation time on our tested benchmark programs. Especially, the simulator-specific PGO (virtual swap) can be applied to the benchmarks to accelerate the simulation speed by a factor of 1.19. As for the hardware-independent PGO, compared with the brute force mechanism (turning on all available compilation flags), which achieves 21\% performance improvement against the non-optimized version, the PGO can achieve 16\% speedup with a factor of 63 less compilation time than the brute force approach. 
\end{abstract}


\section{Introduction}
Quantum circuit simulation mimics quantum systems and their behaviors with classical computers. 
During the early stage of quantum algorithm development, a simulated platform is a more preferred choice than the physical counterpart, and the reasons are threefold. First, quantum software can be developed and validated before the actual hardware is available, shortening the time to market. Second, physical quantum computers are prone to be vulnerable to noise, regarding noisy intermediate-scale quantum~\cite{Preskill_2018}, which prolongs the software development time. Third, a quantum algorithm can be tested and evaluated on different quantum hardware platforms by changing hardware models or simulators to assess a better software/hardware combination.
Thus, major quantum computer vendors offer simulation environments for their customers, in addition to their physical quantum systems, such as Amazon Braket~\cite{braket}, Google Cirq~\cite{cirq}, and IBM Q~\cite{ibm}. The majority of these simulators utilize a full-state quantum circuit simulation methodology. This methodology involves maintaining and updating the quantum states throughout the simulations. The availability of intermediate results during the full-state simulations enhances the ease of debugging. Further details on different methodologies for quantum circuit simulations will be presented in Section~\ref{sec:qs}.

Various companies and research institutes have developed their software and hardware solutions to facilitate quantum computing. 
These software solutions usually support a variety of quantum hardware platforms, especially for the Python-based program development frameworks, Cirq~\cite{cirq}, ProjectQ~\cite{Steiger}, PyQuil~\cite{koch2019introduction}, and Qiskit~\cite{Qiskit}. With these software frameworks, a given quantum program can be run on different hardware without extensive human intervention. This is achieved by using an intermediate representation, such as OpenQASM~\cite{cross2017open}, to represent the input quantum program. This intermediate form is further \emph{transpiled} into the low-level instructions, e.g., gate operations, supported by a physical/virtual quantum processor for execution. 
The transpilation process is usually carried out on the fly before the quantum circuit execution, as the target quantum processor is determined immediately prior to execution. This execution style can be seen in a cloud environment. For instance, when a quantum circuit is submitted into IBM Quantum Cloud, the circuit is queued, validated, and transpiled to the format accepted by a runtime-determined quantum hardware~\cite{parallelizingworkloads}. Under such a circumstance, it is common practice to avoid taking a long time to generate the best transpiled circuit~\cite{blocking_2020}. As a result, it would lose the opportunity for further performance enhancement, especially for hardware-specific optimizations.

This transpilation scheme might work well for physical processors. 
But, it hinders the development of quantum algorithms with virtual processors. For instance, the simulation of a 36-qubit \emph{ising} model on our manycore platform requires approximately ten hours. This simulation employs a \textit{Schrödinger-style} quantum circuit simulator, as will be introduced in Section~\ref{sec:qs}. Given that the simulation is performed in a gate-by-gate manner, the acquisition of simulation results for larger scale problems could span several days\footnote{The time of a quantum circuit simulation, conducted by the gate-by-gate style, grows exponentially with the qubit size. That is, the addition of a single qubit results in a doubling of the simulation time.}. 
Unfortunately, to accelerate the simulation speed, applying all available quantum circuit optimizations during the transpilation process\footnote{For instance, the highest optimization level of the QCOR compiler, which is referred to as an \emph{all-on} configuration, is described in Section~\ref{sec:evaluation}.} would incur a significant amount of compilation time. 
Moreover, this brute force method might not always deliver the best result because of the sequence of applied optimizations\footnote{A to-be-applied optimization may have a negative effect on the program code that has already been optimized by another optimization.}. Actually, finding the best compilation sequence of a given program for a specific hardware platform is an important research topic and has been studied extensively in developing classical compilers~\cite{10.1145/998300.997196}. 

The profile-guided optimization scheme is another approach commonly adopted by classical compilers. This scheme performs program-specific optimizations based on collected runtime performance statistics. 
With the performance profile collected in a prior run, program optimization can be conducted efficiently by focusing on the characteristics of the profile data. 
In the context of quantum circuit simulation, the PGO scheme can be used to generate an improved quantum circuit for accelerating quantum circuit simulations. During the development of quantum algorithms, it is common for a quantum circuit design to undergo multiple simulation runs to verify if the corresponding design meets the expectation.
This is particularly true for variational quantum algorithms, which involve parameters tweaking. PGO can help minimize the time spent on this process by reducing the time of the repeated simulations. For instance, the Quantum Approximate Optimization Algorithm (QAOA) is a popular approach for solving combinational problems. Solving a problem with the QAOA circuit involves multiple simulation runs with varying input data to fine-tune its parameters. Once developed, the QAOA circuit acts as a solver, running repeatedly to provide solutions for given inputs. 
Unfortunately, the classical PGO scheme cannot be applied directly to quantum computing, as will be described below. Thus, a remedy is desired to boost the speed of quantum circuit simulations, facilitating quantum program development. 

\begin{figure}[tbhp!]
  \centering
  \includegraphics[width=1\linewidth]{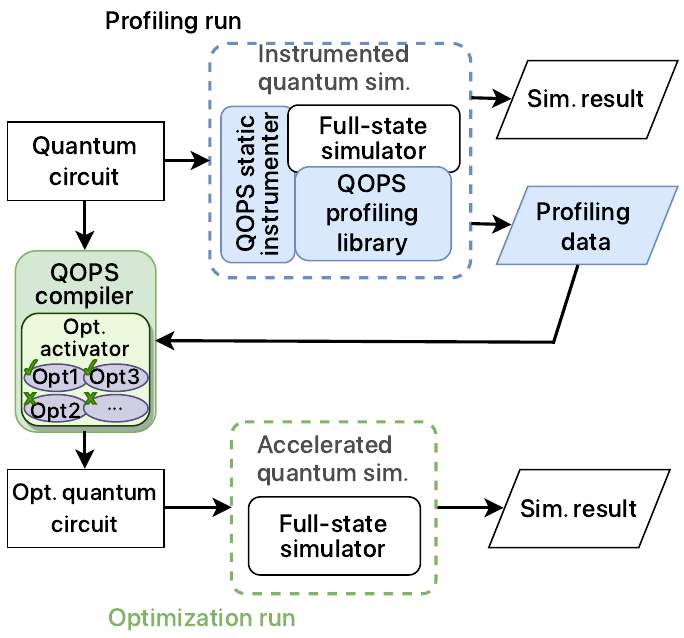}
  \caption{The PGO concept in QOPS for switching on specific optimizations to generate an optimized quantum circuit, based on the profiling data collected in the profiling run.}
  \label{fig.1}
\end{figure}

In this work, we propose a compiler framework \textit{QOPS} to accelerate quantum circuit simulations with the PGO concept, especially for the Schrödinger-style simulators. The major differences between QOPS and the classical PGO scheme are twofold, in terms of the targeting platform and program structure. First, the classical scheme is majorly done on a target program to collect performance statistics during the execution on a physical computer, as will be described in Section~\ref{sec:simulator}. Second, the input data are handled implicitly by program codes in classical computers while in a quantum program, quantum gates are dedicated to operating on specific qubit(s) explicitly. Please refer to Section~\ref{sec:pgo} for further information. In order to deal with the discrepancies, we design and develop a \emph{QOPS profiling library} to collect the performance statistics during simulations via the exposed programming interfaces. 
To alleviate the burden of system developers, a static instrumentation module is added to the \emph{QOPS compiler}. This module helps insert the performance probes offered by the profiling library to an open-source quantum simulator semi-automatically. 
A catch-all module of the PGO for quantum programs is added as an ordinary compiler optimization pass. This optimization activator module turns on the specific program optimization based on the performance profile generated in the prior run of the instrumented simulator. The PGO-based quantum circuit optimization is illustrated in \figurename~\ref{fig.1}.

This work makes the following contributions. 
\begin{enumerate}
    \item A QOPS framework is proposed to perform the profile-guided optimization for quantum circuit simulation acceleration. To the best of our knowledge, this proposal is the first work that boosts the performance of quantum circuit simulations using the PGO concept. Thanks to PGO, a quantum program can be optimized in a more efficient way since the performance profile provides insights to facilitate program- and simulation hardware-specific optimization(s). The PGO scheme avoids the need for turning on all available optimizations at a high compilation-time cost.
    \item A simulation-platform-specific optimization (\emph{virtual swap}) and the two existing optimization algorithms (\emph{single-qubit merge} and \emph{rotation folding}) are included as the profile-guided optimizations. Moreover, the three PGO-enabled optimizations demonstrate the proposed scheme can deliver promising results for quantum circuit simulations. For example, our PGO achieves a 1.44x performance speedup for the QAOA program with 2.3s compilation time, compared with 433s compilation time for the 1.17x speedup by turning on all available optimizations (Section~\ref{sec:pgoresult}).
    \item A QOPS library is proposed to characterize the interactions between quantum software and hardware by the \textit{counter-} and \textit{context-}based profiling schemes. The collected performance data can be used either by the programmers to better understand program behaviors or by a PGO compiler to generate optimized code. 
\end{enumerate}

The QOPS framework has been developed on top of the LLVM-based quantum compiler framework QCOR~\cite{10.1145/3380964}. Encouraging results are obtained in our preliminary experiments. We believe this work pushes forward the study of quantum program optimizations. For instance, the optimizing compiler for the optimization sequence on quantum computers can be done with the performance data collected by the QOPS profiling library. 

In the rest of this paper, the background and related work of existing quantum compilers/simulators and the PGO concept are introduced in \sectionname~\ref{sec:background}, where the links of existing works toward the quantum-based PGO are discussed as well. The overview of the QOPS framework and its key components are presented in \sectionname~\ref{sec:methodology}. The experimental results are provided in \sectionname~\ref{sec:evaluation}. We conclude this work in \sectionname~\ref{sec:conclusion}.

\section{Background and Related Work} \label{sec:background}
This work adds the PGO support for quantum circuits to the LLVM-based quantum compiler for accelerating the full-state quantum circuit simulations. In the following subsections, quantum compilers are first introduced in Section~\ref{sec:backgroundcompiler} with the background information of the QCOR compiler. Next, quantum simulators are introduced in Section~\ref{sec:qs} with emphasis on the full-state simulations, in terms of the simulation approach and simulation schemes (the memory- and storage-based quantum simulators). Lastly, the PGO concept is provided in Section~\ref{sec:pgo}, and the augmentation of the quantum compiler for the PGO support of quantum computing is highlighted. 

\subsection{Quantum Compiler} \label{sec:backgroundcompiler}
A quantum compiler is responsible for converting a high-level description of a quantum algorithm, e.g., Q\#~\cite{Svore_2018}, Scaffold~\cite{10.1145/2597917.2597939}, and QMASM~\cite{7761637}, to lower-level instructions, e.g., cQASM~\cite{khammassi2018cqasm}, Quil~\cite{smith2017practical}, and OpenQASM, which could be accepted by quantum processors. In particular, the conversion involves the decomposition of the high-level program statements into a sequence of quantum gate operations represented by an intermediate representation (IR), e.g., OpenQASM. An IR is used for describing a wide range of quantum operations and for flow control based on the results of measurements. Sometimes, the quantum operations specified by the IR may not be supported by a target quantum processor. In such a case, a low-level circuit translation is required to generate the hardware-specific instructions, i.e., quantum gate operations, for the target processor.

During the conversion process, optimizations are often performed to generate efficient quantum circuits. The optimizations can be categorized into two broad classes, \emph{hardware-independent} and \emph{hardware-dependent} optimizations. The former is usually performed during the translation of an input quantum program to an IR. The latter is often done right before the execution of a quantum circuit on a target quantum processor. Modern quantum compilers, such as IBM's Qiskit~\cite{Qiskit}, ProjectQ~\cite{Steiger}, Cirq~\cite{cirq}, and QCOR~\cite{10.1145/3380964}, are able to perform both classes of optimizations for efficient quantum computing. For example, the above quantum compilers can compile an input quantum program into the OpenQASM format with some hardware-independent optimizations. Then, they transpile the low-level circuit to a desired transpilation target with some hardware-dependent optimizations.

QCOR compiler framework is adopted in this work as it is a C++-based~\cite{nguyen2020extending}, open-source software project, built on top of the LLVM compiler which is one of the most popular compilers for classical computers. More specifically, the QCOR framework incorporates XACC~\cite{McCaskey_2020}, which is a system-level, C++-based infrastructure for hardware-agnostic programming, compiling, and executing a quantum program. With XACC, QCOR is able to support a variety of quantum processors as a multi-target quantum compiler. 
QCOR takes the quantum programs (written in XASM, OpenQASM, or Quil) as its input and translates them into the XACC IR by leveraging the Clang/LLVM infrastructure. The XACC library is then used to transpile the given IR code into the version for a desired quantum processor by the XACC compiler that supports various backend modules targeting different quantum platforms, such as Atos QLM, DWave, IBM, IonQ, and Rigetti QCS. 

QCOR uses a \texttt{PassManager} to handle the quantum code optimizations, where the optimization \textit{passes} are performed at the XACC IR level via the XACC \texttt{IRTransformation} functions. 
The \texttt{PassManager} is responsible for invoking the XACC pass(es) functions to perform the desired optimization(s). In addition to leveraging the circuit optimizations offered by XACC, QCOR has made available XACC \texttt{IRTransformation} plugins to support the optimizations brought by the third-party quantum circuit optimization libraries, such as the rotation folding in the Staq library~\cite{Amy_2020} and the OCaml-based verified optimizer for quantum circuits (VOQC) optimizers~\cite{Hietala_2021}.

In this work, the above optimization pass design is adopted to add the PGO support to QCOR, using our implemented optimization activator. Both hardware-independent optimizations in QCOR and our developed hardware-dependent optimization can be selectively turned on to generate an optimal quantum circuit. 
While state-of-the-art quantum compilers implement many typical quantum circuit optimizations, they do not take advantage of this profile-data guideded optimization concept. These typical quantum compilers are orthogonal to our proposed PGO scheme. Therefore, this PGO design can be added to these quantum compilers for more efficient quantum circuit optimizations. Further information on our proposed design is provided in Section~\ref{sec:compiler}.

\subsection{Quantum Simulator} \label{sec:qs}
A quantum simulator can emulate the behaviors of quantum systems on classical computers to avoid the need for a physical quantum computer. In particular, researchers can develop, test, and verify quantum algorithms on simulation platforms. As an emerging technology, various quantum simulators have been developed to serve different purposes. A quantum circuit simulation can be done on a cluster system to handle a large-qubit quantum circuit. The quantum simulator \emph{Qulacs}~\cite{Suzuki_2021} developed by Fujitsu can handle up to 48-qubit systems on the Fujitsu's Fugaku supercomputer~\cite{9336127}. On the other hand, the simulation can also be done on a commodity computer (with a lower total cost of ownership). For example, the storage-based quantum circuit simulation techniques have been developed to simulate the quantum circuits with several tens of qubits, using multithreading CPUs/GPUs~\cite{NVMeSSD, 10.1145/3524059.3532375}. 

Two major types of methodologies are usually adopted to emulate the behaviors of quantum systems: \textit{amplitude sampling} and  \textit{Schrödinger-style} approach~\cite{Liu_2021, li2018quantum, pednault2020paretoefficient}. The former method is computation intensive and suitable for simulating a shallow circuit, and it is often adopted for quantum chip validation. The latter approach is memory intensive and suitable for a deep circuit and is good for evaluating the efficiency of a quantum algorithm. 
In particular, the Schrödinger-style simulation approach~\cite{9407212} is often adopted to emulate the states of qubits affected by the input quantum circuits (so it is also known as full-state simulation). This kind of simulation approach is able to produce intermediate results (states) and can generate the final probability distribution result. Given an $n$-qubit quantum circuit system, the Schrödinger-style approach keeps the $2^{n}$ complex states in the memory subsystem on the underlying simulation hardware, and the states are updated gate-by-gate during the simulation~\cite{Jones_2019, markov2018quantum, De_Raedt_2019, De_Raedt_2007, pednault2020paretoefficient, li2018quantum, Liu_2021}. 

In order to overcome the limitation brought by buffering the qubit states during the full-state simulations, an emerging trend is to incorporate the secondary memory to accommodate those states, in addition to the main memory. As the secondary memories (e.g., hard disk drives and solid state drives, SSDs) have a relatively lower cost and a higher capacity, they pave a way for scaling quantum circuit simulations with larger qubits~\cite{NVMeSSD, 10.1145/3524059.3532375}. 
In this work, the storage-based simulator is adopted to demonstrate the effectiveness of our PGO concept~\cite{NVMeSSD}. 
This simulator proposes a data representation method to partition the $n$-qubits into segments, including \emph{file segment}, \emph{middle segment}, and \emph{chunk segment}. The $2^n$ quantum states of an $n$-qubit system are initially stored in SSDs and indexed by these segments. These quantum states are accessed by their indexes when they need to be updated to reflect the effects of an emulated quantum gate. For example, given an $n$-qubit index, the first few bits belong to the file segment, representing a quantum state file, which is accessed by standard file I/O operations. The remaining bits of the middle and chunk segment are the offset for accessing the complex amplitudes of a quantum state. In this case, it is beneficial to update for those quantum gates whose target qubits fall within the same chunk segment. This is because the quantum states within a chunk segment would be buffered in the primary memory, thanks to the file system caching mechanism in operating systems. Hence, a chunk-segment-aware state updating mechanism can improve the simulation performance significantly. A different accessing pattern that jumps around quantum state files would result in poor simulation performance. More information on the qubit indexing design is available in~\cite[Figure 4]{NVMeSSD}.

It is important to note that our proposed PGO scheme can profile the simulation time of quantum gates on different qubits. This profiling information guides the \emph{virtual swap} optimization to increase the opportunities that the accessing to the quantum states is concentrated on a buffered chunk segment for better simulation performance. An example of the optimization is illustrated in \figurename~\ref{fig:virtual-swap}. 
The figure shows that a hotspot is presented on the $q_{0}$ for handling the Pauli gates, where the state data are spread across different files\footnote{On the top of \figurename~\ref{fig:virtual-swap}, the quantum states with the qubit patterns, `1******' and `0******', will be accessed. These states are stored on different files. Note that as a small qubit size is used in this example, the middle segment is omitted for the qubit indexing. }. Therefore, a simulator-specific optimization is desired to eliminate the file I/O operations via a swap operation for the Pauli operations, thereby shortening the simulation time. More about the \emph{virtual swap} optimization~\cite{ICCETW} is presented in Section~\ref{sec:optimization}. 

\subsection{Profile-Guided Optimization} \label{sec:pgo}
PGO is a widely adopted technique of traditional compilers to produce the optimized code for classical computers, where the optimizations are guided by the performance profiling data of the same code obtained in a previous execution run. As the performance profile captures the runtime behaviors, traditional compilers can put emphasis on the optimizations for the \emph{hot code regions}. This can make the best use of the compilation time on the hot codes for producing an efficient machine executable. The two popular classical compilers, GCC and LLVM, support the PGO feature to produce efficient code. In particular, these compilers collect the instruction-level performance statistics, such as the total count (or frequency) of the executed instructions, or the frequently-executed code paths. Later, the classical compilers will spend time on the most frequently executed code for further optimizations. For example, the function inlining can be adopted for those function calls on the frequently executed code paths to eliminate the function invocation overheads~\cite{PGO, MSPGO}. 

There are performance profiling tools for quantum circuits (e.g., qprof~\cite{10.1145/3529398} which imitates the default profiler \texttt{gprof} of GCC). Unfortunately, their primary goal is to provide the performance data for revealing the high-level runtime behaviors, such as caller-callee relationships and the call counts. These performance data offer hints for quantum programmers to optimize their code, which would require extensive human intervention. 
Furthermore, those quantum compilers that build on top of the classical compilers do not support the PGO feature, which adds the burden of optimizing quantum circuits to the quantum programmers. 
Therefore, this work aims to augment the PGO support for quantum programs on the LLVM-based compiler, QCOR. 
Furthermore, our developed methodology serves as a plugin to the LLVM framework to work in parallel with the original PGO infrastructure in LLVM.

The traditional programs running on classical computers share very different characteristics from quantum programs. 
First, more sophisticated program structures are present in classical computing, such as having many functions with complex caller-callee relationships, than the structures of quantum programs. The classical PGO methodology is to reveal such relationships for further optimizations.
Second, the execution of a classical program on a classical computer is a sequence of executed instructions, i.e., one-dimensional data. On the contrary, the execution of a quantum circuit on a quantum platform involves cross-qubit data manipulations, e.g., by two-qubit gates, which can be considered as two-dimensional data. Hence, a profiling data format is devised to characterize the runtime behaviors of a quantum circuit during the simulation. More about the collected profiling data is presented in Section~\ref{sec:profiling}. 

\section{Methodology}\label{sec:methodology}

\begin{figure*}[tbhp!]
  \centering
  \includegraphics[width=0.68\linewidth]{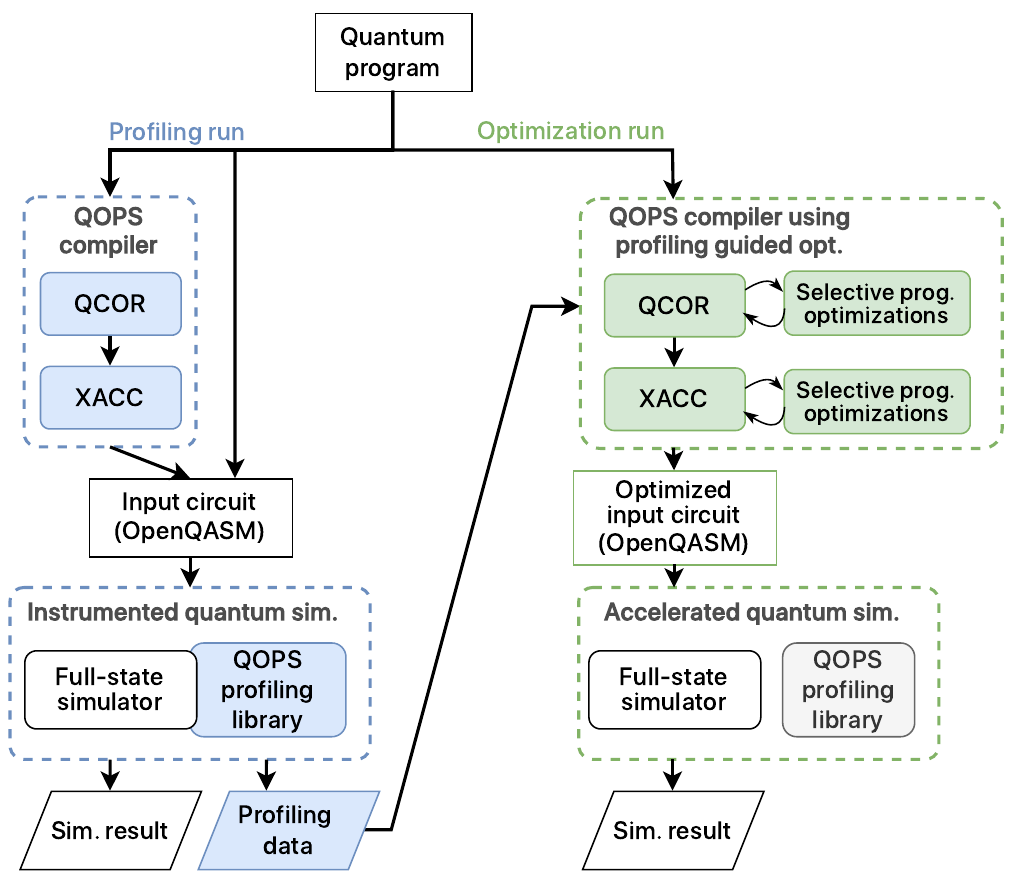}
  \caption{The workflow for the QOPS framework, where it involves a \textit{profiling run} to collect the performance statistics by the instrumented quantum simulator, and an \textit{optimization run} to accelerate the quantum circuit simulation with the input circuit that is compiled with the optimizations guided by the profiling data of the profiling run.}
  \label{fig:workflow}
\end{figure*}

The QOPS framework operates in two different simulation runs: a \textit{profiling} run and an \textit{optimization} run, as illustrated in \figurename~\ref{fig:workflow}. In the profiling run, a given full-state quantum simulator is instrumented first. The simulator takes a quantum circuit as an input for quantum circuit simulation, where the input can be generated by QOPS compiler or by some other quantum compiler\footnote{The two different paths, from the quantum program to the input circuit in the profiling run, are used to depict the two different situations.}. The instrumented simulator collects runtime performance statistics by our QOPS profiling library during the simulation. The generated performance profile is fed into the QOPS compiler to guide further optimizations.

In the optimization run, the input quantum program is first compiled by the QOPS compiler which performs the code optimization(s) selectively based on the profiling data collected in the profiling run. Next, the QOPS compiler generates the optimized, converted circuit in the format accepted by the quantum simulator, e.g., OpenQASM. Finally, the original, unmodified version of the quantum simulator is used to do the accelerated simulation. Note that in this run, the QOPS profiling library has no effect and does not interfere with the quantum circuit simulation. 

\subsection{QOPS Compiler} \label{sec:compiler}

The compiler operates differently in the two different runs. In the profiling run, it acts as an ordinary quantum compiler that can convert a given quantum program into the corresponding circuit. Note that since the QOPS compiler is built upon the QCOR compiler, leveraging the XACC library to generate target circuits, it can target different quantum computers, as described in Section~\ref{sec:backgroundcompiler}. In our experiments, we switch on/off all the optimization flags supported by QCOR to estimate the performance with/without the optimizations, in terms of compilation time and simulation time of an input quantum program. In addition, the QOPS compiler is able to turn an ordinary quantum simulator into a PGO-enabled version (described in Section~\ref{sec:simulator}) with the QOPS profiling library (introduced in Section~\ref{sec:profiling}).

\begin{figure*}[bhtp!]
    \centering
    \includegraphics[width=0.9\linewidth]{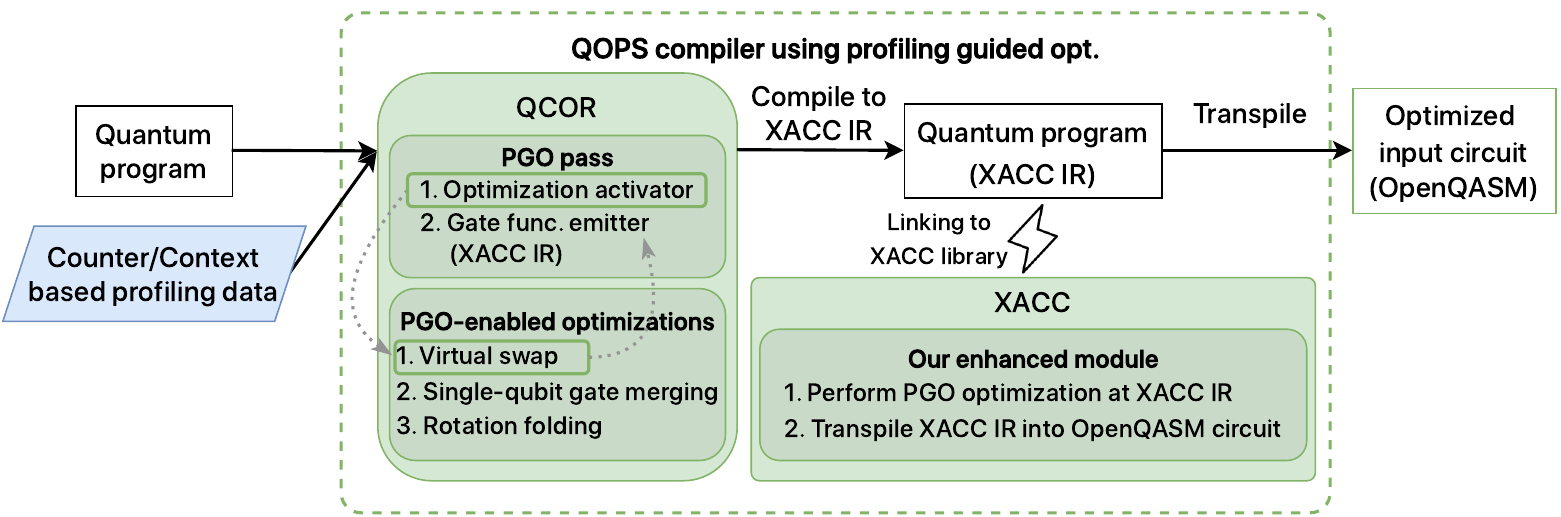}
    \caption{A zoom-in view of the QOPS compiler using profiling guided optimization shown on the top-right corner of Fig.~\ref{fig:workflow}.}
    \label{fig:qopsinternal}
\end{figure*}

In the optimization run, the compiler takes the same quantum program and the profiling data generated in the profiling run as its inputs, as shown in \figurename~\ref{fig:qopsinternal}. It then selectively switches on the optimization based on the collected performance data, which is enabled by compiling the program with the compilation flag \texttt{-opt-pass} to activate our developed \texttt{PGO} pass. After performing the code optimization, it generates the corresponding format of the circuit for the targeting quantum processor, e.g., OpenQASM.

\textbf{\emph{PGO design considerations.}} There are two important issues to be considered for the PGO scheme. First, the mapping of the profiling data and the to-be-applied optimization(s) should be determined so as to take advantage of the PGO scheme. Second, the sequence of performing the to-be-applied optimizations is important since one optimization can cancel the effect done by another optimization. 
Determining the mapping and the optimization sequence is a complex problem since it requires searching in the huge space comprising the optimization flags offered by the compiler and the performance metrics to characterize the simulated quantum program. Fortunately, the research on the selection of the compilation optimizations and the optimization sequence for a given computer program has been conducted extensively in the literature for classical computers, e.g., an automatic compiler optimizations selection framework for embedded applications~\cite{DBLP:conf/icess/HungTLC09}. The existing efforts can be leveraged in the future.

In the current version, in order to validate the effectiveness of the PGO concept, the values of the performance counters are used as the indicators for applying the optimizations selectively. Furthermore, an optimization flag is turned on at a time to avoid the negative effect introduced by the subsequent optimization(s). It is interesting to note that the counter-based, single-optimization approach works effectively for some quantum programs as these programs have relatively simpler program structures than classical programs. Note that the counter data can serve as an important basis for further designing a sophisticated mapping algorithm. More about the supported performance counters are presented in Section~\ref{sec:profiling}.

\textbf{\emph{PGO compilation flow.}} 
The developed PGO pass is a catch-all entry for all profile-guided optimizations included in QOPS. This optimization pass examines the profiling data and activates the specific optimization when a pre-defined condition is satisfied, as will be introduced in Section~\ref{sec:optimization}. In \figurename~\ref{fig:qopsinternal}, the virtual swap optimization is activated by the optimization activator in the PGO pass. This optimization inserts the virtual swap gates into the input quantum program by issuing the QCOR APIs to emit the corresponding swap gates into the program in the XACC IR form. The quantum program in the XACC IR form is then transpiled into the OpenQASM circuit via the XACC library. That is, the library converts the high-level gates represented by the XACC IR into the actual quantum gates supported by the target quantum processor. 

\textbf{\emph{PGO-enabled optimizations.}} 
The three optimizations have been included in the PGO scheme, including our developed \textit{virtual swap} algorithm~\cite{ICCETW} and the two QCOR built-in compiler optimizations, single-qubit gate merging and rotation folding. 
The purpose of the inclusion of the newly developed and the two existing optimizations is to demonstrate the flexibility and efficiency of the PGO concept. It can be used to add a new optimization pass or incorporate an existing one. This is possible as long as the relationships between the profiling data and the optimizations are uncovered. 
For the sake of completeness, the adopted PGO optimizations are introduced in Section~\ref{sec:optimization}. 

\subsection{Instrumented Quantum Simulator for PGO} \label{sec:simulator}

An ordinary, full-state quantum simulator can be turned into a PGO-enabled simulator with the QOPS profiling library. It can be done automatically by the QOPS compiler or manually by the developers. 
This design is made possible due to the observation that some of the full-state quantum circuit simulations are performed with the Schrödinger-style algorithm. In this case, during the gate-by-gate simulation, the designated function is responsible for mimicking the behavior/effect of a to-be-simulated quantum gate. The performance data can be collected by injecting the performance probes, e.g., the APIs exposed by the QOPS profiling library, to those designated functions, and the performance data will be recorded during the profiling run. 

Alternatively, a simulator can use one single function to do the gate simulations. For instance, it could be implemented by using a big case-switch structure to dispatch to a proper code segment for simulating a quantum gate. In this regard, the parameter used to dispatch the gate simulation code should be further analyzed to instrument proper probes to the respective gate-simulation code segments. The concept of the instrumented simulator using the QOPS profiling library to collect the performance data is illustrated in the bottom-left corner of Fig.~\ref{fig:workflow}. 

The collected performance data can be further used either by a PGO-enabled compiler, e.g., our QOPS compiler, or by the developers of the simulator. The proposed QOPS compiler uses the profiling data to switch on the compiler optimizations judiciously. For example, as will be introduced in Section~\ref{sec:profiling}, the counts of the simulated gates recorded in the profiling data, e.g., using the counters for the simulated gates, are used as an indicator to trigger the optimization pass(es) for the frequently-simulated gates. On the other hand, the simulation counts of the gates are helpful for the simulator developers to further optimize the implementations of the frequently simulated gates to accelerate the simulation speed. 

\subsection{QOPS Profiling Library} \label{sec:profiling}
The QOPS profiling library provides the means to collect the performance data during the quantum circuit simulation, such as which qubits are used by the gates, the elapsed time of simulated gates, and the sequence of the simulated gates. Two types of profiling techniques are implemented in the library, \emph{counter-based} and \emph{context-based} profiling, which are introduced in the following paragraphs, respectively. The profiling overhead is shown in Section~\ref{sec:sqs}. The small profiling overhead will benefit the development time for the large-scale quantum circuit simulation. When the simulation runs for days, the profiling data will be dumped.  

\begin{figure}[hbtp!]
    \centering
    \includegraphics[width=1\linewidth]{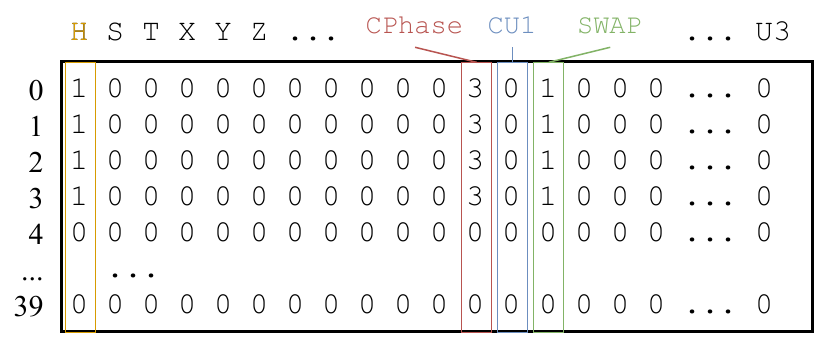}
    \captionof{lstlisting}{The counter-based profiling result (zero fields are ignored). The gate numbers of different gate types on each qubit are reported.}
    \label{lst:counter_profile_data}
\end{figure}

\textbf{\emph{The counter-based profiling}} provides statistical information on the simulated gates. For example, our current implementation supports a total of thirty-three counters, such as H, S, T, X, and U gates, each of which reflects the count of a simulated gate in the simulation. The support of a new counter for a new gate is simple by adding a new variable to keep the count of a new gate in our library. As the counter-based profiling simply calculates the occurrence number of each type of simulated gate and dumps the counter data into the file, it incurs negligent overhead to the quantum circuit simulation itself. An example of the output for the counter-based profiling is illustrated in the Listing~\ref{lst:counter_profile_data}. In particular, each column shows the occurrence number of a gate (the first column is H, the second one is S, the third one is T, ..., the 11th one is CPhase, the 12th one is CU1, and the 13th one is SWAP, etc.) and different rows represent the data on a distinct qubit. A line number indicates a qubit number, e.g., 0 for $q_0$. We support 40-qubit simulation currently and it can be extended to more qubits easily. With these qubit-wise-simulation-count data, it would be helpful for the QOPS compiler to switch on some gate-specific optimization in the optimization run. For example, rotation folding optimization is useful for handling the merging T gate phases situation.

\begin{figure}[hbtp!]
    \centering
    \includegraphics[width=1\linewidth]{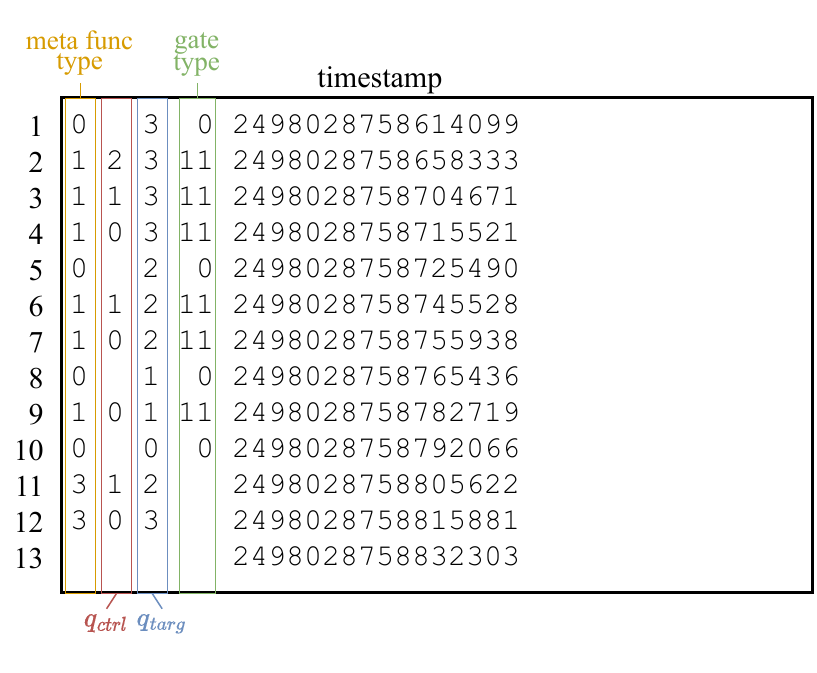}
    \captionof{lstlisting}{The context-based profiling result for recording the sequence of simulated quantum gates with the timing information.}
    \label{lst:context_profile_data}
\end{figure}

\textbf{\emph{The context-based profiling}} tracks the sequence of the simulated quantum gates. In the current implementation, the library uses a tuple record to track each simulated gate. Take Listing~\ref{lst:context_profile_data} as an example.
The first element of the tuple, i.e., the first column, is the meta function type to identify whether the gate is a single-qubit gate (denoted as 0), a two-qubit gate (denoted as 1), or a SWAP gate (denoted as 3). If it is a single-qubit gate, the following element represents the target qubit. Otherwise, the two following elements are used for a two-qubit gate to record control/target qubits, respectively. In detail, the second element of the first row, i.e., the number 3, means the $q_3$ is the target qubit of the first gate, H; the second row identifies $q_2$ controls $q_3$ by a CPhase gate. The next element specifies the gate type used to tell the simulator which operation should be performed. For instance, we use 0 to represent an H gate and 11 for a CPhase gate, respectively. The SWAP gate is an exception in that it does not have an element for gate type because there is only one type in the SWAP meta function. The last element of the tuple is the timestamp of the simulated gate. 
The gate sequence information is able to further increase the opportunity for PGO since it has the ability to discover the combination of simulated gates, which are able to be merged to shorten the simulation time. For example, the sequence of two consecutive \textit{CNOT} and \textit{CNOT} gates can be eliminated, and the sequence of the rotation gates, such as S and T, can be merged to save the simulation time. Additionally, context-based profiling is able to reveal the time sequence of the simulated quantum gates and this is helpful for the debugging purpose. 
Moreover, the profiling data can be transferred to a timeline diagram for the simulated quantum circuit, as will be presented in Section~\ref{sec:scalingresult}. 

The profiling data are collected by instrumenting the performance probes into a given quantum simulator, as described in Section~\ref{sec:simulator}. In this version, as proof of the PGO concept, the instrumentation is done statically on the source code of a given, full-state simulator by the QOPS compiler. 
Specifically, the optimization pass infrastructure offered by Clang/LLVM is adopted to insert the performance probes into the source code of the simulator. The static code instrumentation requires prior knowledge of the function names for the simulation of quantum gates if each quantum gate uses a dedicated function to do the gate simulation. In fact, a Clang plugin is implemented with a class inherited \texttt{ModulePass} that requires the implementation of the  \texttt{runOnModule} function. Besides, in order to get high-precision time information, a POSIX API \texttt{clock\_gettime} is called to report the timestamp of a to-be-simulated gate for context-based profiling. 
It is interesting to note that a dynamic instrumentation scheme, e.g., \texttt{ftrace}, can be adopted to avoid the need for the source code to perform the PGO-based simulation acceleration, and it is considered to be one of our future working directions. 

\subsection{Implemented Profile-Guided Optimizations} \label{sec:optimization}
The three profile-guided optimizations are included in the current QOPS implementation,  \emph{virtual swap}, \emph{rotation folding}, and \emph{single-qubit gate merging}. 
In the optimization run, our developed optimization activator uses profiling data to determine whether a quantum circuit optimization should be enabled based on its decision rules. For instance, the activator evaluates the consecutive single-qubit gate counter, e.g., for the Pauli gate and Hadamard gate. When the count of the consecutive single-qubit gates exceeds a certain threshold, e.g., five, the single-qubit gate merging optimization is activated to fuse the consecutive single-qubit gates.

\begin{figure*}[htbp!]
  \centering
  \includegraphics[width=1\linewidth]{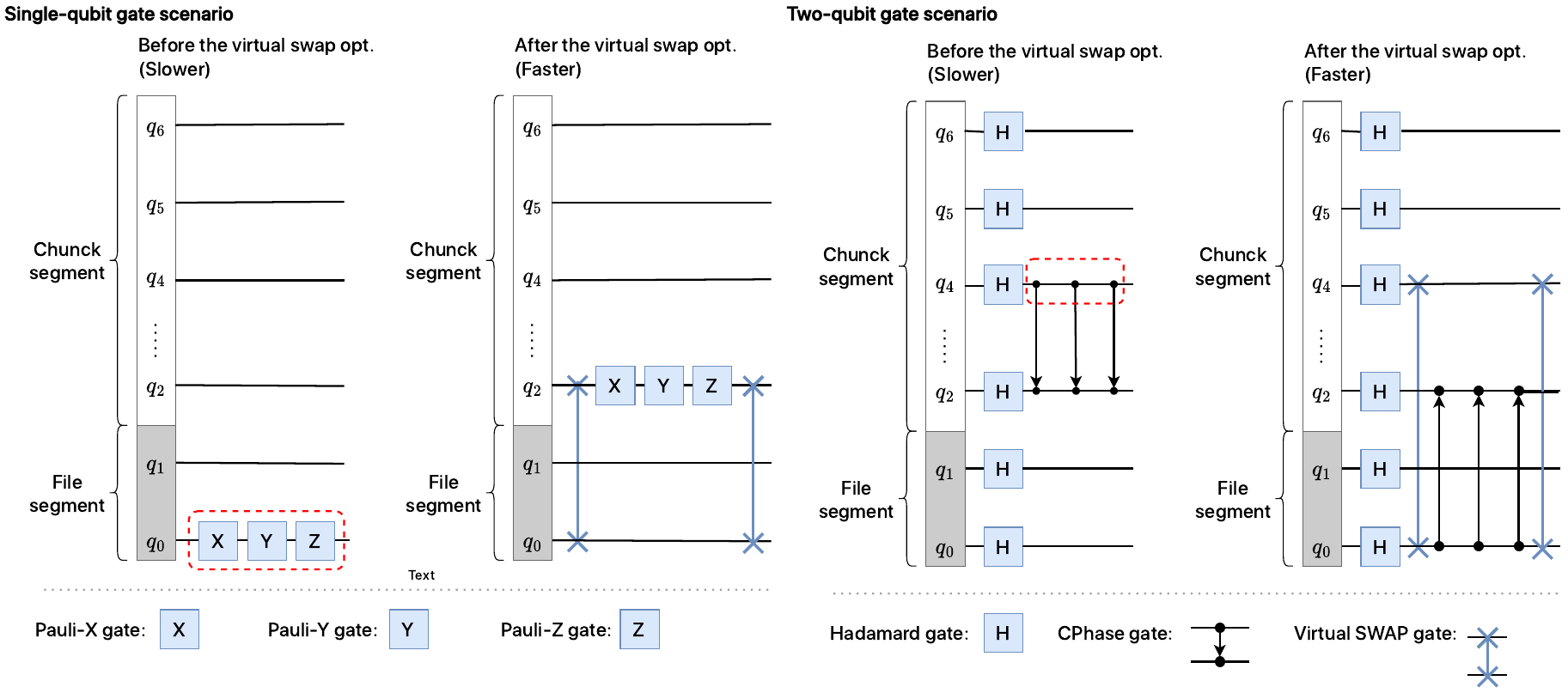}
  \caption{Illustration of the virtual swap optimization for the single-qubit (between $q_0$ \& $q_{2}$) and two-qubit gates (between $q_0$ \& $q_{4}$). The execution time for quantum gates are different in distinct segments (file segment and chunk segment).}
  \label{fig:virtual-swap}
  \end{figure*}

\textbf{\emph{Virtual swap}}~\cite{ICCETW} inserts the virtual swap gates based on statistical timestamp profiling results to perform the optimization for the storage-based simulator introduced in Section~\ref{sec:qs}. \figurename~\ref{fig:virtual-swap} illustrates the concept of the virtual swap optimization for the single- and two-qubit gates situations. On the left of the figure, to simulate the behaviors of the X, Y, and Z gates on $q_0$, the simulation threads have to manipulate the quantum states across different state files. This incurs file accesses, and accessing the quantum states from files is longer than the virtual-swap optimized version of manipulating the state data in the chunks that are buffered in the memory. On the other hand, as for the two-qubit gate scenario, the fastest arrangement is when the control qubit is resident in the file segment and the target qubit is at the chunk segment. This arrangement helps enjoy the higher parallel file I/O performance when accessing the states of the qubit pair. To activate this optimization in the optimization run, the QOPS compiler examines the values of the two-qubit gate counters, e.g., CPhase and CNOT gates, belonging to the chunk segment, and the compiler performs the optimization as long as the values of these counters above a pre-defined threshold. 

\textbf{\emph{Rotation folding}}, an existing optimization pass in QCOR, is utilized to reduce the number of total quantum gates. The primary approach involves combining rotated phases using commutation rules applicable to any Clifford and Pauli gates. While T-count optimization may not be critical for NISQ programs, 
reducing T and other phase gates have been shown to frequently lead to significant CNOT reductions, which are more costly in a NISQ setting~\cite{Amy_2020}. The Listings~\ref{list:before_rf} and \ref{list:after_rf}, two equivalent circuit snippets, illustrate the impact of rotation-folding optimization in reducing the total number of gates. 
The compiler follows the commutative rules~\cite{zhang2019optimizing, Amy_2020} for the T gates and the CNOT gates to convert the code fragment in Listing~\ref{list:before_rf} to Listing~\ref{list:after_rf}. 
In QCOR, the rotation folding optimization encompasses Staq's~\cite{Amy_2020} rotation folding optimizer, part of a C++-based library for optimizing quantum circuits in the OpenQASM format. 
To activate rotation folding as a PGO, the QOPS compiler utilizes T-gate frequency as an indicator. When the ratio of the T-gate count to the total gate count exceeds a threshold (e.g., 10\%), the rotation-folding optimization is enabled to explore opportunities for converting the input quantum program into a simplified version.
\Suppressnumber
\begin{table}[t]
\centering
\footnotesize
\begin{minipage}[t]{.4\linewidth}
\begin{code}[caption={The OpenQASM code before the rotation folding opt. for ham15\_high. CX is CNOT gate.},label=list:before_rf,escapechar=|,escapeinside=||,firstnumber=12,mathescape=true]
// ...|\Reactivatenumber{13}|
t q[9]; // T gate|\Reactivatenumber{14}|
t q[19];|\Reactivatenumber{15}|
h q[19]; // H gate|\Reactivatenumber{16}|
CX q[5], q[9]; |\Reactivatenumber{17}|
t q[5];|\Reactivatenumber{18}|
tdg q[9]; // T$^\dagger$ gate|\Reactivatenumber{19}|
CX q[5], q[9];|\Suppressnumber|
// ...|\Reactivatenumber{163}|
t q[9];|\Suppressnumber|
// ...|\Reactivatenumber{166}|
CX q[5], q[9];|\Reactivatenumber{167}|
t q[5]; |\Reactivatenumber{168}|
tdg q[9];|\Suppressnumber|
// ...|\Suppressnumber|
\end{code}
\end{minipage}
\qquad\quad
\begin{minipage}[t]{.4\linewidth}
\begin{code}[caption={The OpenQASM code after the rotation folding opt. for ham15\_high. CX is CNOT gate.},label=list:after_rf,escapechar=|,firstnumber=12,mathescape=true]
// ...|\Reactivatenumber{13}|
t q[19];|\Reactivatenumber{14}|
h q[19];|\Suppressnumber|
// ...|\Reactivatenumber{123}|
s q[9]; // S gate|\Suppressnumber|
// ...|\Reactivatenumber{126}|
CX q[5], q[9];|\Reactivatenumber{127}|
s q[5];|\Reactivatenumber{128}|
sdg q[9]; // S$^\dagger$ gate|\Suppressnumber|
// ...
\end{code}
\end{minipage}
\end{table}

\textbf{\emph{Single-qubit gate merging}} is another QCOR optimization that is used to merge the sequence of consecutive single-qubit gates. According to the Z-Y decomposition~\cite{nielsen_chuang_2010} for a single qubit gate method, all single qubit gates can be decomposed and be represented by a unitary gate and rotation gates. Therefore, consecutive single-qubit gates can be merged into the rotation gates and reduce the number of total quantum gates. For example, when the number of consecutive single-qubit gates is larger than a threshold (e.g., five), the single-qubit-gate-merging optimization is switched on looking for the opportunity to convert the input quantum program into the simplified version.

 It should be noted that the above quantum circuit optimizations serve as examples to showcase how existing quantum circuit optimizations can be transformed into PGO-enabled optimizations. This transformation is achieved by briefing the effects of the three quantum circuit optimizations and incorporating the profiling data, as introduced in \sectionname~\ref{sec:profiling}. 
 With the proposed PGO scheme, unnecessary compilation time can be avoided when some quantum circuit optimizations are not applicable to a given quantum circuit program. For instance, not all quantum circuits consist of consecutive single-qubit gates targeting the same qubit. A notable counterexample is the Quantum Fourier Transform (QFT), and its structure is illustrated in \figurename~\ref{fig:visualization}. The QFT program includes H gates, making the single-qubit gate optimization inapplicable to QFT. Consequently, activating this optimization would be inefficient and a waste of time. 
 This inefficiency becomes more pronounced as the number of non-applicable optimizations increases, especially when dealing with larger quantum circuit programs. 
This situation arises when a full-fledged quantum compiler, equipped with many quantum circuit optimizations, activates all available optimizations without considering their applicability, leading to significant time consumption. Therefore, integrating the PGO scheme into the compiler presents an efficient solution for quantum circuit compilation and simulation. 


\section{Results} \label{sec:evaluation}

The QOPS framework is built on top of the open-source compiler projects: QCOR (commit hash baedb52) and XACC (commit hash 8e3b96f). The nine benchmark programs~\cite{benchmarks} are selected from the QCOR compiler (four reversible logic synthesis benchmarks consisting of \emph{``adder''}, hidden weighted bit function \emph{``hbt6''}, \emph{``grover''} algorithm, and hamming coding function \emph{``ham15''}), and the popular quantum algorithms (the variational quantum eigensolver \emph{``vqe''}, the quantum approximate optimization algorithm \emph{``qaoa''}, the \emph{``STO-3G''} algorithm in chemistry, the quantum Fourier transform \emph{``qft''}, and the \emph{``ising''} model). The experiments done in this work are performed on the manycore computing platform, where the hardware and software specifications are listed in Table~\ref{tab:platform}. Each reported performance datum is the average number of ten runs.

In order to demonstrate that the QOPS is practical and can actually improve the simulation performance, the parallel, storage-based quantum simulator~\cite{NVMeSSD} is adopted and instrumented by the QOPS compiler automatically via the static instrumentation on the source code of the simulator.  Section~\ref{sec:sqs} introduces this instrumentation detail, as well as the overhead incurred by the instrumented performance code for the profiling run. Section~\ref{sec:pgoresult} presents the advantages provided by the PGO scheme, where the compilation optimizations can be turned on selectively to achieve similar performance results that are delivered by the \emph{all-on} configuration of switching on all compilation flags\footnote{In fact, all available optimizations within the highest optimization level (\texttt{O2}) in QCOR are included in this configuration, including the rotation-folding, single-qubit-gate-merging, circuit-optimizer, and two-qubit-block-merging optimizations.}. 
This is considered as the comparison between the selective scheme and the \emph{all-on} scheme, using the performance metrics of the compilation time and the delivered speedup. Section~\ref{sec:scalingresult} further shows the scaling of the simulation provided by the virtual swap optimization. Section~\ref{sec:debugperf} presents how the context-based profiling data can be used to help the debugging of the quantum circuit design and its performance analysis. Section~\ref{sec:largescale} showcases the benefit brought by the PGO scheme with a real-world use case of multiple simulation runs for a given quantum circuit.

\begin{table}[tbp!h]
    \caption{Hardware and software specifications of the experiments.}
    \label{tab:platform}
    \begin{tabular}{p{1.3cm}p{6.4cm}}
      \toprule
      \textbf{Attribute} & \textbf{Description} \\
      \midrule
      CPU & AMD Ryzen Threadripper PRO 3995WX 64-Cores @ 2.7GHz (64C/128T) \\
      \specialrule{0em}{1pt}{1pt}
      RAM & HX436C18FB3K2/64, 256GB (8x32GB) DDR4 3600MHz \\
      \specialrule{0em}{1pt}{1pt}
      Storage & Kingston SFYRD/2000G, 16TB (each 2TB, PCIe 4.0 x4)\\
      \specialrule{0em}{1pt}{1pt}
      OS  & Ubuntu 20.04 LTS (kernel version 5.15.0-58-generic) \\
      \specialrule{0em}{1pt}{1pt}
      Compiler & QCOR (commit hash baedb52; extending from LLVM ver. 12.0.0 by ORNL-QCI), XACC (commit hash 8e3b96f), and Staq library ver. 2.0 \\
      Threading & OpenMP library ver. 4.5 \\
      \specialrule{0em}{1pt}{1pt}
      \bottomrule
    \end{tabular}
  \end{table}
  
\subsection{Instrumenting a Storage-based Quantum Simulator} \label{sec:sqs}
The storage-based quantum simulator~\cite{NVMeSSD} (introduced in Section~\ref{sec:qs}) is instrumented to validate the implementation of the proposed QOPS framework, in terms of the performance data collection and the profile-guided optimization. The performance probe instrumentation is done similarly to an ordinary compilation process for a quantum simulator. That is, the QOPS compiler takes the source code of the storage-based simulator as an input with an extra compiler flag \texttt{-profile-mode}, and it generates the machine executable for the classical computer, i.e., the AMD Ryzen processor in our experiment. 

Moreover, there are two other inputs are needed for the automatic static instrumentation: 1) the names of the gate-simulation functions and 2) the mapping of the simulated gates and the parameters to the gate-simulation functions. Taking the storage-based simulator as an example, it has one single gate-simulation function, called \texttt{single\_gate}, to mimic the behaviors of different quantum gates, and the value of its argument is used to specify the gate to be simulated. With the mapping of the type of a quantum gate and the gate-simulation function with the value of the argument, the QOPS compiler is able to track the gate count by counter-based data and the executed gates by context-based data. Based on the setting to \texttt{-profile-mode}, the \texttt{insertCounterProbe} or \texttt{insertContextProbe} handler will be invoked to insert the counter-/context-based probes (exposed by the QOPS profiling library) right before the call site of the gate-simulation function. 

The performance overheads for the two profiling schemes are evaluated on the nine benchmark programs listed in Table~\ref{tab:allonresult}. 
These overheads, which are introduced in the profiling run to collect the profiling data needed by the PGO scheme, are extra time added to the original simulation time, excluding the compilation time. Slowdown ratios are used to quantify the overhead for counter- and context-based profiling, $S_{counter}$ and $S_{context}$, as shown in Table~\ref{tab:traceoverhead}. 
On average, counter-based profiling exhibits a 3.1\% runtime overhead, while context-based profiling incurs a 5.7\% runtime overhead. Both are the result of weighted averages based on the gate count of benchmarks. The reasons for the low profiling overheads are twofold. First, the simulator is instrumented based on the flow presented in Section~\ref{sec:profiling}, and the instrumented simulator is saved as the LLVM binary-level format, \emph{bitcode}, for further process. Our analysis shows that during this lower-level bitcode conversion, the LLVM compiler performs certain optimizations on the instrumented code\footnote{The two simulator variants with and without the code instrumentation are compiled with the \texttt{-O3} setting of LLVM to generate bitcodes.}, generating more efficient instructions. These extra optimizations lead to a lower overhead in the profiling run. Notably, for the ham15 program, the counter-based profiling is a little bit faster than the non-modified version. 
Second, the experimental hardware platform is able to cache the context-based profiling data in the file system buffer, hiding the file IO latencies. Based on our analysis, the nine benchmarks do not drain the hardware resources, a total of 256 GB of main memory, and fast SSDs. This is also the reason that the two profiling schemes have a similar overhead.
It is important to note that the simulation results are valid in the profiling run. This means that the real cost of the PGO scheme is the above profiling overhead and the simulation time is not wasted, which is an important factor for a large quantum circuit program requiring a longer simulation time. 
Further experiments are required to explore the profiling overhead for the simulations with larger qubits.

Overall, this design facilitates the retargeting process, i.e., users only need to update the above two inputs when targeting another full-state quantum simulator. 
Thanks to the LLVM compiler pass infrastructure, our QOPS compiler is able to do the profiling of quantum circuit simulations on a classical computer and optimize a quantum circuit running on a simulated platform. The following subsection presents the performance enhancement brought by the PGO feature. 

\subsection{PGO Performance Impact} \label{sec:pgoresult}
Tables~\ref{tab:allonresult} and~\ref{tab:threeoptresult} present the performance of the \emph{all-on} and \emph{selective} configurations, respectively. Two performance metrics, compilation times and speedups, are compared across these configurations. 
Table~\ref{tab:allonresult} lists the delivered performance brought by turning on all of the available QCOR built-in compilation flags to observe the full potential of the \textit{all-on} configuration\footnote{Each case is averaged from 10 runs of a specific benchmark.}. 
The performance metrics include the number of the generated gates before $N_{ori}$ and after the optimization $N_{opt}$, the simulation time before $T_{ori}$ and after optimization $T_{opt}$, the speedup $\mathit{S}$, and the compilation time for the \emph{all-on} configuration $T_{cmp}$. The baseline performance offered by the QCOR compiler is characterized by $N_{opt}$, $T_{opt}$, $\mathit{S}$, and $T_{cmp}$ in the table. 
The results show that six out of nine programs have relatively shorter simulation time after applying the all-on compilation configuration (1.25x speedup on average), especially for the Ising model (1.73x). 
Those programs with a larger number of quantum gates tend to require a longer compilation time. In this case, turning on all available compilation optimizations seem not to be a good idea since it might take too long for compilation, e.g., $\sim$4 hours\footnote{Based on our analysis, the \texttt{circuit-optimizer} optimization in QCOR contains a collection of simple pattern-matching-based circuit optimizations, and it often dominates the overall compilation time.}. This hampers the iterative quantum algorithm development process. For instance, the compilation time required by the two programs, sto3g and ham15, exceeds our timeout threshold: fifteen minutes, and hence, their performance data are excluded from the table. 

Table~\ref{tab:threeoptresult} provides the performance delivered by each of the three optimizations presented in Section~\ref{sec:optimization}, demonstrating the potential of the PGO-enabled optimizations. 
The performance delivered by each optimization is characterized by the compilation time $T_{cmp}$ (in seconds) and the achieved speedup $\mathit{S}$ after applying the optimization, where the weighted averages of $T_{cmp}$ and $\mathit{S}$ are based on the gate count of the nine benchmarks mentioned above. 
For example, the \emph{ising\_model\_10} program achieves the highest performance speedup of 1.72 with the single-qubit merging optimization, at a compilation time cost of 1.6 seconds. This significant performance boost is attributed to a remarkable number of quantum gates being merged, resulting in an approximately 40\% reduction in the total quantum gates. Our observations are summarized as follows. 
\begin{enumerate}
    \item Different programs exhibit their distinct characteristics and there is no single optimization that can achieve the best performance across all nine programs. That is, the single-qubit gate merging is the best for the \emph{ising} model, the rotation folding is good for \emph{grover}, \emph{ham15}, and \emph{adder}, and the virtual swap is the best for \emph{hwb6}. 
    \item The simulator-specific optimization\footnote{This virtual swap optimization is not included as the built-in optimizations in QCOR.} delivers an average speedup of 1.19, which is comparable to that of the all-on configuration (1.25x speedup), with 65 times less compilation time. This result demonstrates the importance of simulator-specific optimization. 
    \item One can obtain a significant performance improvement by applying a single optimization, e.g., rotation folding, without turning on all optimizations. 
    For instance, excluding the outlier \emph{ising} and the timeout circuits \emph{sto3g} and \emph{ham15}, the average speedup for the all-on configuration in Table~\ref{tab:allonresult} is 1.21. The rotation folding optimization yields a speedup of 1.16, but with the advantage of requiring 63 times less compilation time, as indicated in Table~\ref{tab:threeoptresult}.
    \item It is an interesting case that the quantum circuit pattern of \emph{qft\_n18} is hard to optimize, even for the simulator-specific optimization. In such a case, it would take a longer compilation time for the all-on scheme ($\sim$40s), but it delivers no performance improvement. On the other hand, the cost of the PGO-enabled optimization is far less ($<$2s). This is another evidence of activating the optimizations judiciously can help facilitate the quantum circuit development process. 
\end{enumerate}

The all-on scheme signifies the best efforts that can be achieved by traditional quantum compilers, which do not rely on the feedback information to guide the optimizations. The above experimental results showcase that PGO can obtain similar performance speedups $\mathit{S}$ with far less compilation time $T_{cmp}$ (Table~\ref{tab:threeoptresult}), in contrast to the all-on configuration (Table~\ref{tab:allonresult}). The listed observations further justify the importance and benefits of the PGO scheme. It is important to note that the PGO design is not bounded by QCOR, and can be implemented on other quantum compilers for more efficient quantum circuit compilations.

\begin{table}[hbt]
  \centering
  \caption{The performance results delivered by turning on all available compilation optimizations on QCOR (the \textit{all-on} configuration). The performance metrics include $N_{ori}$ and $N_{opt}$ for original and optimized gate numbers, $T_{ori}$ and $T_{opt}$ for simulation runtimes of original and optimized quantum circuits, $\mathit{S}$ for the speedup number achieved by the optimized quantum circuit and $T_{cmp}$ for compilation time of a PGO. The unit of time is in seconds.} \label{tab:allonresult}
  \begin{tabular}{rrrrrrr}
  \toprule
    & \begin{tabular}[c]{@{}l@{}}$N_{ori}$ \\\end{tabular} 
    & \begin{tabular}[c]{@{}l@{}}$N_{opt}$ \\\end{tabular} 
    & \begin{tabular}[c]{@{}l@{}}$T_{ori}$ \\\end{tabular} 
    & \begin{tabular}[c]{@{}l@{}}$T_{opt}$ \\\end{tabular} 
    & \begin{tabular}[c]{@{}l@{}}$S$\\\end{tabular} 
    & \begin{tabular}[c]{@{}l@{}}$T_{cmp}$\\ \end{tabular} \\ \midrule
  hwb6 & 257 & 230 & 1.2 & 1.0 & 1.14 & 2.5\\ \hline
  h4\_adapt\_vqe & 1,908 & 1,558 & 7.9 & 6.3 & 1.26 & 246.4\\ \hline
  h4\_qaoa & 2,060 & 1,826 & 8.8 & 7.6 & 1.17 & 433.6 \\ \hline
  grover\_5 & 831 & 634 & 3.3 & 2.4 & 1.35 & 21.1 \\ \hline
  ising\_model\_10  & 480 & 284 & 2.1 & 1.2 & 1.73 & 3.6 \\ \hline
  NH\_frz\_P\_sto3g & 7,289 & $T_{out}$ & 27.2 & $T_{out}$ & $T_{out}$ & $T_{out}$ \\ \hline
  qft\_n18 & 801 & 801 & 3.3 & 3.3 & 1.00 & 40.5 \\ \hline
  ham15\_high & 5,308 & $T_{out}$ & 20.3 & $T_{out}$ & $T_{out}$ & $T_{out}$ \\ \hline
  adder\_8 & 934 & 735 & 3.5 & 2.7 & 1.31 & 40.1 \\ \bottomrule
  \end{tabular}
  \end{table}

    \begin{table}[hbt]
        \centering
        \caption{The performance results delivered by turning on a single PGO-enabled optimization; \textbf{Swap} for virtual swap opt., \textbf{Rotation} for rotation folding opt., and \textbf{Merge} for single-qubit gate merging opt. The performance metrics include $\mathit{S}$ for the speedup number and $T_{cmp}$ for compilation time of a PGO-enabled optimization.} \label{tab:threeoptresult}
        \begin{tabular}{rrrrrrr}
            \toprule
            & \multicolumn{2}{l}{\textbf{Swap}} & \multicolumn{2}{l}{\textbf{Rotation}} & \multicolumn{2}{l}{\textbf{Merge}} \\ \midrule
            & \begin{tabular}[c]{@{}l@{}}$S$\\ \end{tabular}     
            & \begin{tabular}[c]{@{}l@{}}$T_{cmp}$\\\end{tabular}   
            & \begin{tabular}[c]{@{}l@{}}$S$\\ \end{tabular} 
            & \begin{tabular}[c]{@{}l@{}}$T_{cmp}$\\\end{tabular}      
            & \begin{tabular}[c]{@{}l@{}}$S$\\ \end{tabular}       
            & \begin{tabular}[c]{@{}l@{}}$T_{cmp}$\\\end{tabular}      \\ \hline
          {hwb6} & \textbf{1.55}        & 1.5       & 1.13      & 1.5        & 1.02       & 1.5     \\ \hline
          {h4\_adapt\_vqe} & \textbf{1.43}        & 2.1       & 1.24      & 2.2        & 1.10       & 2.2     \\ \hline
          {h4\_qaoa} & \textbf{1.44}        & 2.2       & 1.15      & 2.3        & 1.08       & 2.3    \\ \hline
          {grover\_5} & 1.18        & 1.6       & \textbf{1.26}      & 1.6        & 1.12       & 1.7     \\ \hline
          {ising\_model\_10} &  1.12       & 1.5       & 1.21      & 1.5        & \textbf{1.72}       & 1.6     \\ \hline          
          {NH\_frz\_P\_ sto3g} & \textbf{1.18}        & 4.6       & 1.00      & 4.8        & 1.10       & 5.1       \\ \hline
          {qft\_n18} &  \textbf{1.09}       & 1.7       & 1.00      & 1.7        & 1.00       & 1.7     \\ \hline
          {ham15\_high} & 1.04        & 3.3       & \textbf{1.33}      & 3.5        & 1.06       & 3.5     \\ \hline
          {adder\_8} & 1.04        & 1.7       & \textbf{1.28}      & 1.7        & 1.03       & 1.7     \\ \hline \hline
          {\textit{Average}} & 1.19        & 3.3       & 1.16      & 3.4        & 1.09       & 3.5 \\
          \bottomrule
          \end{tabular}
    \end{table}

    \begin{table}[hbt]
      \centering
      \caption{The overheads added on the quantum circuit simulation time during the profiling run, where $S_{context}$ and $S_{counter}$ are for slowdown ratios incurred by context- and counter-based profiling, respectively.} \label{tab:traceoverhead}
      \begin{tabular}{rrrrrrrrr}
      \toprule
          & \begin{tabular}[c]{@{}l@{}}$S_{context}$\\ \end{tabular}
          & \begin{tabular}[c]{@{}l@{}}$S_{counter}$\\ \end{tabular} \\ \midrule
      hwb6 & 1.08 & 1.00\\ \hline
      h4\_adapt\_vqe & 1.16 & 1.14 \\ \hline
      h4\_qaoa  & 1.06 & 1.04 \\ \hline
      grover\_5 & 1.06 & 1.03 \\ \hline
      ising\_model\_10  & 1.13 & 1.17\\ \hline
      NH\_frz\_P\_sto3g & 1.07 & 1.05\\ \hline
      qft\_n18 & 1.00 & 1.03 \\ \hline
      ham15\_high & 1.00 & 0.95 \\ \hline 
      adder\_8 & 1.03 & 1.03 \\ \hline \hline
      {\textit{Average}} & 1.05        & 1.03\\\bottomrule
      \end{tabular}
      \end{table}

    \subsection{Scaling Quantum Circuit Simulation with the Virtual Swap Opt.} \label{sec:scalingresult}
    Scalability is an important feature for quantum circuit simulations. This experiment is used to evaluate if the simulator-specific optimization has positive effects for larger qubit sizes. To this end, we select the \emph{qft\_n18} circuit from the tested benchmark programs and the reasons are threefold. First, it is widely adopted by famous quantum algorithms, such as Shor's algorithm. Second, ordinary hardware-independent optimizations have a negligent effect on its delivered performance, as shown in Table~\ref{tab:allonresult}. Third, it has a moderate gate number and does not require too huge simulation time when considering a larger qubit size. 

Table~\ref{tab:virtual swap number} lists the simulation time before and after the optimization of the virtual swap method and the related speed-up numbers, with the qubit sizes ranging from 20 to 34. Overall, the simulator-specific optimization can achieve up to 13\% performance improvement across different qubit sizes. Note that some of the qubit sizes do not perform well (e.g., 1\% improvement for the 20-qubit experiment) since the same configuration of the chunk size and file size is applied to the storage-based quantum circuit simulator for all of the qubit sizes experiment to observe the performance improvement achieved by the virtual swap. Further performance enhancement can be achieved by tweaking the chunk/file sizes, which is left for our future work. 

It is also important to note that when performing the simulation for the 34-qubit QFT circuit, the simulation time is about ten times larger than that of the 32-qubit version. This is because the 256 GB system memory is not large enough to accommodate the required quantum states, and the state data are partially kept in the solid state drives. Accessing the data from SSDs results in a prolonged simulation time. The experimental results are important evidence, showing that the PGO-enabled optimization is effective even when the simulator behaves differently, i.e., memory-intensive (qubit size $\le$ 32) or storage-intensive (qubit size $\ge$ 34).    

        \begin{table}[bthp!]
            \centering
            \caption{The performance achieved by the virtual swap optimization of the Quantum Fourier Transform circuit for different qubit sizes. The unit of time is in seconds.}\label{tab:virtual swap number}
            \begin{tabular}{crrr}
            \toprule
               \emph{Qubit} & \begin{tabular}[c]{@{}l@{}}$T_{ori}$\\\end{tabular} & \begin{tabular}[c]{@{}l@{}}$T_{opt}$\\\end{tabular} & \begin{tabular}[c]{@{}l@{}}$S$\\\end{tabular} \\ \hline
            20 & 1,519                                                    & 1,508                                                 & 1.01                                                 \\ \hline
            22 & 1,926                                                    & 1,792                                                 & 1.07                                                 \\ \hline
            24 & 2,382                                                    & 2,181                                                 & 1.09                                                \\ \hline
            26 & 2,801                                                    & 2,585                                                 & 1.08                                                \\ \hline
            28 & 3,339                                                    & 3,047                                                 & 1.10                                                \\ \hline
            30 & 3,942                                                    & 3,635                                                 & 1.08                                                \\ \hline
            32 & 4,592                                                    & 4,218                                                 & 1.09                                                \\ \hline
            34 & 56,678                                                    & 50,363                                                 & 1.13                                                \\  \bottomrule
            \end{tabular}
            \end{table}

\subsection{Performance Debug and Analysis} \label{sec:debugperf}
\begin{figure*}[tbhp!]
  \centering
  \includegraphics[width=1\linewidth]{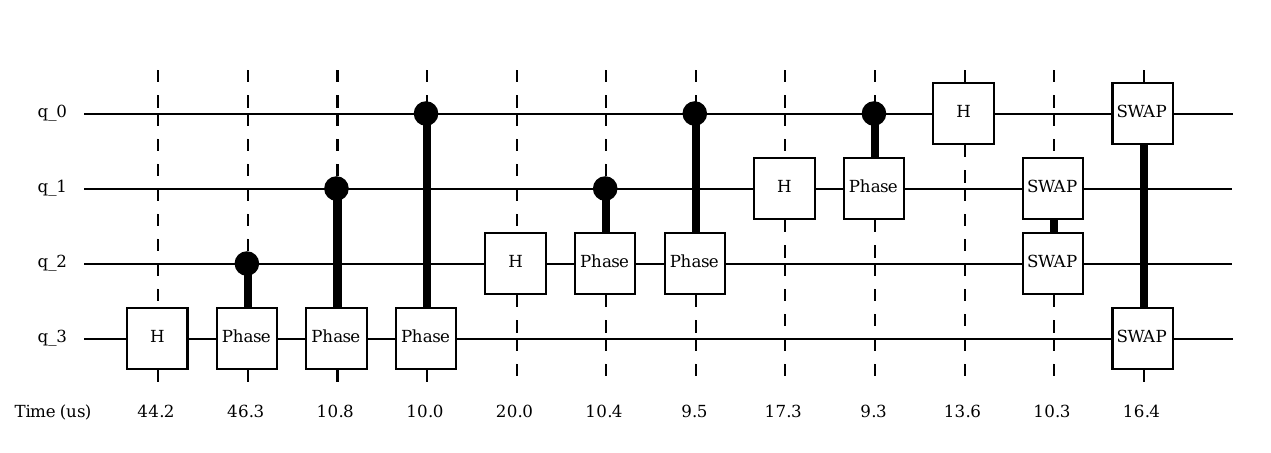}
  \caption{Quantum circuit visualization for performance debugging, converted by our developed tool from the context-based profiling data in Listing~\ref{lst:context_profile_data}.}
  \label{fig:visualization}
\end{figure*}

In addition to supporting the profile-guided optimization, the counter- and context-based profiling data shown in Listings~\ref{lst:counter_profile_data} and~\ref{lst:context_profile_data} can also be used for performance debugging and analysis by compiler optimization developers and quantum programmers. The former can be further turned into a histogram depicting the counts of different gates for each qubit, where different circuit optimizations can be performed on a per-qubit(s) basis. The latter contains the timing information of the simulated quantum circuit, which is useful for programmers to examine the time-costly gate operations. We have built a tool to convert the collected context-based profiling data into a visualized form. \figurename~\ref{fig:visualization} gives an example timing information of a 4-qubit QFT program by revealing the elapsed time of each simulated gate operation. The first H gate and the following three CPhase gates whose target is $q_3$ have a higher simulation time because of the low-speed segment. The rest of the gates have a relatively shorter simulation time. 
With the help of such a timeline diagram, the programmers are able to perform higher-level optimizations by replacing the time-consuming gates with lower-cost gates while satisfying the high-level algorithm logic. 

\subsection{Handling Multiple Simulation Runs of a Quantum Circuit} 
\label{sec:largescale}
Multiple simulation runs of a given quantum circuit are required during the development of certain quantum algorithms. A notable example is Quantum Approximate Optimization Algorithm, which is a popular approach for solving combinational optimization problems. During the development of a QAOA algorithm, a critical part is to establish the ansatz for the corresponding QAOA circuit of a target problem. To find the optimal states (parameters) for the quantum circuit, a hybrid iterative method is adopted with a classical optimizer for finding the proper ground states\footnote{In particular, angle parameters $\gamma$ and $\beta$ are iteratively updated by the optimizer.} based on the results obtained by each run of the quantum circuit simulation. Besides, in order to improve the quality\footnote{Higher approximation ratio.} of the offered solutions, a layer parameter setting varialbe $\rho$ is often adopted to apply the the QAOA cirucit $\rho$ times~\cite{farhi2014quantum}. 
This repetition introduces complexity for the parameter finding and increases the depth of the quantum circuit, resulting in a longer simulation time. 
After the training process, the quantum circuit is executed with these parameters to provide the solution to the problem. As a solver to the target problem, the quantum circuit would be exercised repeatedly to find good results (by taking several simulation runs to build the expectation value for the established QAOA circuit), or to provide the problem solutions (when the solver responds to the service requests made by multiple clients). 

In light of the above scenario, a QAOA circuit is built for solving the Max-Cut problem~\cite{farhi2014quantum} to showcase the advantages of the PGO scheme. A 36-qubit QAOA circuit is constructed\footnote{The QUBO problem and the Max-Cut problem can be formulated by each other and both solved by the QAOA algorithm~\cite{rehfeldt2022faster}.}~\cite{qaoa} to handle the Max-Cut problem of a graph with 36 nodes. The layer parameter setting variable $\rho$ ranges from one to four to reflect the situation for achieving superior results~\cite{qaoainqaoa}. 
The following experiments perform 20 simulation runs for each $\rho$ setting, representing that the QAOA solver runs multiple times to generate the results. Based on the profiling data, the RZ gate counter has a relatively larger value, activating the single-qubit gate merging optimization. Table~\ref{tab:large scale} lists the performance data before and after the optimization, along with the achieved speedup for applying the single-qubit gate merging optimization. The simulation time indicates the accumulated simulation time required by the multiple-run quantum circuit simulation for generating the results.
Similar to the performance data reported in Table~\ref{tab:threeoptresult}, significant performance gains are achieved in this case study by turning on the PGO activated optimization, e.g., 1.74 times faster than the original version for the \emph{qaoa\_ansatz\_p4} configuration. This speedup increases as the $\rho$ value grows.

It is important to note that the compilation time of the \emph{all-on} configuration is 17\% longer than that of the PGO configuration due to a smaller circuit size of 260 quantum gates when $\rho$ is four. Considering a larger $\rho$ value, i.e., $\rho=20$, the quantum circuit contains 1,184 quantum gates. This circuit takes 1.8 seconds to compile with the PGO configuration, while it requires 317 seconds to process with the \texttt{circuit optimizer} optimization within the \emph{all-on} configuration since this specific optimization contains optimizations for different gate patterns as described in Section~\ref{sec:pgoresult}.
The situation of prolonged compilation time will worsen when considering a full-blown quantum compiler, which implements a large number of quantum circuit optimizations. Activating all available optimizations on such a comprehensive quantum compiler would require a significant amount of compilation time. Hence, a profiling-guided scheme is a better alternative to judiciously activate useful optimizations, accelerating quantum circuit simulations.

\begin{table}[hbt]
  \centering
  \caption{The performance data of the 36-qubit QAOA circuit with different $\rho$ levels for solving the Max-Cut problem. The performance metrics include $N_{ori}$ and $N_{opt}$ for representing the total simulated gates before and after the quantum circuit optimization, $T_{ori}$ and $T_{opt}$ for indicating the total simulation time before and after the optimization applied, and $\mathit{S}$ for the performance speedup achieved by the applied optimization. The unit of time is in days.}\label{tab:large scale}
  \begin{tabular}{rrrrrr}
    \toprule
                        & $N_{ori}$  & $N_{opt}$  & $T_{ori}$    & $T_{opt}$    & S    \\\hline
  qaoa\_ansatz\_p1 & 1,840  & 1,680  & 2.22  & 2.12  & 1.05 \\\hline
  qaoa\_ansatz\_p2 & 2,960 & 2,080  & 3.06 & 2.47  & 1.24 \\\hline
  qaoa\_ansatz\_p3 & 4,060 & 2,520 & 4.17 & 3.16  & 1.32 \\\hline
  qaoa\_ansatz\_p4 & 5,200 & 2,860 & 5.54 & 3.18 & 1.74 \\\bottomrule 
  \end{tabular}
  \end{table}

\section{Conclusion} \label{sec:conclusion}
The QOPS framework is proposed to boost the performance of quantum circuit simulations. 
The QOPS framework includes the PGO-enabled compiler and the profiling library. The PGO-enabled compiler is built on top of the QCOR compiler. The compiler is able to statically instrument the full-state quantum simulator~\cite{NVMeSSD} with the performance probes for collecting the performance data via the profiling library. The collected performance profile is used to judiciously activate certain quantum circuit optimizations during quantum circuit compilations. 
In addition, the collected data can be used for debugging purposes and for performance enhancement by the PGO-enabled compiler. Encouraging results are obtained on the three PGO-enabled optimizations on the nine benchmarks. The results of the virtual swap optimization demonstrate that PGO is a good way to perform simulation-hardware-specific optimizations. 
In comparison to the all-on configuration that yields a speedup of 1.25, the virtual swap optimization achieves an average speedup of 1.19 with 65 times less compilation time. The inclusion of the two existing circuit optimizations further demonstrates that the QOPS framework can take advantage of existing efforts for simulation acceleration. The promising results obtained for the 36-qubit QAOA suggest that the proposed framework is useful for the real-world use case. We believe that the QOPS framework paves the way towards feedback-guided optimizations in quantum computing.

\section*{Acknowledgement}

This work was supported in part by National Science and Technology Council, Taiwan, under Grants 113-2119-M-002-024 and 111-2221-E-006 -116 -MY3.

\balance
\bibliographystyle{IEEEtran}
\bibliography{paper.bib}
\end{document}